# Title

Generalised linear models for prognosis and intervention: Theory, practice, and implications for machine learning

# Authors

Kellyn F Arnold[1,2], Vinny Davies[3], Marc de Kamps[1,4], Peter WG Tennant[1,2,5], John Mbotwa[1,2], *Mark S Gilthorpe[1,2,5]

[1]Leeds Institute for Data Analytics, University of Leeds, Leeds, UK; [2]School of Medicine, University of Leeds, Leeds, UK; [3]School of Computing Science, University of Glasgow, Glasgow, UK; [4]School of Computing, University of Leeds, Leeds, UK; [5]The Alan Turing Institute, London, UK

*_Corresponding author:_ Mark S Gilthorpe. Leeds Institute for Data Analytics, University of Leeds, Level 11 Worsley Building, Clarendon Way, Leeds, LS2 9NL, UK. Email: M.S.Gilthorpe@leeds.ac.uk. Tel: +44 (0) 113 343 1913.

# Word count

3908

# Abstract

Prediction and causal explanation are fundamentally distinct tasks of data analysis. In health applications, this difference can be understood in terms of the difference between prognosis (prediction) and prevention/treatment (causal explanation). Nevertheless, these two concepts are often conflated in practice. We use the framework of generalised linear models (GLMs) to illustrate that predictive and causal queries require distinct processes for their application and subsequent interpretation of results. In particular, we identify five primary ways in which GLMs for prediction differ from GLMs for causal inference: (1) The covariates that should be considered for inclusion in (and possibly exclusion from) the model; (2) How a suitable set of covariates to include in the model is determined; (3) Which covariates are ultimately selected, and what functional form (i.e. parameterisation) they take; (4) How the model is evaluated; and (5) How the model is interpreted. We outline some of the potential consequences of failing to acknowledge and respect these differences, and additionally consider the implications for machine learning (ML) methods. We then conclude with three recommendations which we hope will help ensure that both prediction and causal modelling are used appropriately and to greatest effect in health research.

# 1 Introduction

Prediction and causal explanation are fundamentally distinct tasks of data analysis.[1] A thorough discussion of this distinction is given by Shmueli,[2] yet the analytical implications are poorly recognised in much of health research,[3] for which the distinction can be understood in terms of the difference between *prognosis* (prediction) and *prevention/treatment* (causal explanation).

Whilst many of the same techniques (e.g. linear models) can be applied to both predictive and causal queries, they require distinct processes for their application and subsequent interpretation of results. This is perhaps most easily demonstrated in the context of generalised linear models (GLMs), but has applicability to other modelling methodologies, including machine learning (ML). For this reason, we attempt here to simply and concisely illustrate the key differences between prediction and causal inference in the context of generalised linear models (GLMs), to outline the potential consequences of failing to acknowledge and respect these differences, and to provide recommendations which might enable prediction and causal modelling to be used effectively in health research.

# 2 A brief introduction to generalised linear models (GLMs) and historical sources of confusion

Multiple (linear) regression models estimate the expected value $E(*)$ of a single variate $Y$ (the 'dependent' or 'outcome' variable) from a linear combination of a set of observed covariates $X_1, \ldots, X_n$ (the 'independent' or 'explanatory' variables, or simply 'predictors'), as in:

$$E(Y) = \hat{\beta}_0 + \hat{\beta}_1 X_1 + \cdots + \hat{\beta}_n X_n$$

GLMs offer greater flexibility to accommodate a wider range of outcome distributions by allowing a *function* of the outcome (i.e. the 'link function' $f(*)$) to vary linearly with respect to the covariates, as in:

$$f(E(Y)) = \hat{\beta}_0 + \hat{\beta}_1 X_1 + \cdots + \hat{\beta}_n X_n \quad \text{(Eq.1)}$$

The coefficients $\hat{\beta}_0, \hat{\beta}_1, \ldots, \hat{\beta}_n$ for a given GLM are typically obtained via a statistical process known as 'maximum likelihood estimation', which determines the values which make the observed data 'most likely'.[4] Although GLMs are theoretically simple to understand and implement, estimating their parameters without the aid of a computing device quickly becomes intractable as the number of covariates grows.

The emergence of programmable desktop computers in the 1980s and 1990s therefore facilitated a revolution in data analytics, since it became possible to perform both swiftly and automatically the complex matrix inversions required for generalised linear modelling. However, the routine application of generalised linear modelling that became established and entrenched was unwittingly predicated on *prediction*, rather than *causal explanation*.

Standard GLMs are agnostic to the causal structure of the data to which they are fitted. The process of fitting a GLM makes no assumptions about causality, nor does it enable any conclusions about causality to be drawn without further strong assumptions. Pearl's work on graphical causal models (often in the form of directed acyclic graphs – DAGs) provides a formal framework for causal inference using GLMs by explicating the assumptions required to interpret individual coefficients as causal effects.[5-7] However, utilisation of this framework (and, indeed, recognition of its existence) has been limited in health research.[8, 9]

As a result (and despite consistent reminders that 'correlation does not imply causation'), it remains common practice to endow the estimated coefficients for individual covariates with causal meaning, often on the basis of 'statistical significance'. This may be done explicitly or implicitly, as in a recent (though by no means unique) high-profile study which found a significant *association* between active commuting and lower risk of cardiovascular disease but then used this as the basis for recommending initiatives that support active commuting.[10]

Persistent confusion has also been created by much of the language used to describe the relationships between a dependent variable and its 'predictors', which often serves to conflate correlational relationships with causal ones. Perhaps the most notorious example of this is the term 'risk factor', which has both associational and causal connotations across different contexts.[3, 11]

These factors have combined to produce ambiguity about how GLMs for prediction differ from GLMs for causal inference, often resulting in the conflation of two distinct concepts.

## 3    GLMs for prediction and causal inference

Models for prediction are concerned with optimally deriving the likely *value* (or *risk*) of an outcome (i.e. $Y$ in Eq.1) given information from one or more 'predictors', a key task of risk prediction and prognosis. In contrast, models for causal inference are concerned with optimally deriving the likely *change* in an outcome (i.e. $\hat{\beta}_i$ for $1 \leq i \leq n$ in Eq.1) due to (potentially hypothetical) change in a *particular covariate* (i.e. $X_i$), a key task of prevention and treatment. Models for prediction and causal inference are thus fundamentally distinct in terms of their purpose and utility, and methods optimised for one cannot be assumed to be optimal for the other.

GLMs for prediction and causal inference differ with respect to:

1. The covariates that should be considered for inclusion in (and possibly exclusion from) the model;
2. How a suitable set of covariates to include in the model is determined;
3. Which covariates are ultimately selected, and what functional form (i.e. parameterisation) they take;
4. How the model is evaluated; and
5. How the model is interpreted.

To illustrate these differences, we use for context a recent study by Pabinger et al.[12] published in the Lancet Haematology which concerns venous thromboembolism (VTE), a common complication of cancer in which a blood clot forms in a deep vein and then becomes lodged in the lungs. We consider how two research questions – one predictive, one causal – might be addressed using logistic regression (i.e. GLMs with the 'logit' link function) in this context. This is then followed by a more general discussion regarding the implications these differences have for the application and interpretation of GLMs in health research, and subsequent implications for ML.

### 3.1 Prediction modelling

The ultimate utility of a prediction model lies in its ability to accurately predict the outcome of interest. Such information may be used to *anticipate* the outcome – either to simply prepare for its occurrence or to inform a subsequent intervention that attempts to alter it.

In our clinical context, for instance, a prediction model for VTE in cancer patients could be used to identify individuals at heightened risk of VTE solely so that they can be more carefully monitored in hospital, or so that they can receive treatment with low-molecular-weight heparin[13] in order to *reduce* the risk that has been predicted.

The *prediction* research question (RQ1, and that which is addressed by Pabinger et al.[12]) can thus be framed as:

*Which cancer patients are most (or least) likely to develop VTE?*

#### 3.1.1 Which covariates should be considered for inclusion in the model?

Variables which are hypothesised to be useful 'predictors' of the outcome should be identified; these are variables which are likely to be *associated* with the outcome, though not necessarily directly causally related to it. As an example, Pabinger et al.[12] consider D-dimer concentration as a possible covariate. D-dimer is a protein that is present in the blood only after the coagulation system has been activated, and thus a marker for the *existence* of a blood clot rather than a *cause* of it.

Practical considerations often restrict the set of variables that are considered. For example, where a specific dataset has already been chosen, only variables which appear in this dataset are considered for inclusion. Variables that are easy to measure and/or record are also preferred, in order to improve the practical utility of the final model.

### *3.1.2 How are covariates selected for inclusion in the model, which covariates are ultimately selected, and how are they parameterised?*

Methods for narrowing down the set of 'candidate' covariates are generally automatic and/or algorithmic in nature (e.g. best subsets regression, forwards/backwards stepwise or change-in-estimate procedures[14]) and operate within a restricted range of the infinite potential parametric space. These methods evaluate different possible covariate subsets and parameterisations according to some specified criteria in order to arrive at a suitable model. Pabinger et al.,[12] for instance, implement the least absolute shrinkage and selection operator (LASSO[15]) combined with a backwards selection algorithm to narrow down their twenty 'candidate' covariates to just two – tumour site and (log-transformed) D-dimer concentration.

The covariates that are ultimately selected are those which, *as a group*, are deemed to efficiently maximise the amount of joint information relative to the outcome. Selecting the 'optimal' subset of covariates typically involves a trade-off between 'explaining' the maximum amount of variation in the outcome and creating a parsimonious model that is likely to fit other similar datasets. Additional complexity – in the form of more covariates and more complex parameterisations – is likely to increase the predictive capabilities of the model, but at the expense of external validity. This trade-off is made explicit in many of the criteria used for subset selection, including adjusted $R^2$ and penalised likelihood-based measures such as Akaike's Information Criterion (AIC) and Bayesian Information Criterion (BIC).[4]

### *3.1.3 How is the model evaluated?*

The model is assessed via statistical evaluation of the overall model, with little focus on the specific choice of model covariates and limited reference (if any) to exogenous theoretical constructs informed by the context of the model. Examples of common 'goodness-of-fit' criteria include root mean squared error of residuals, (adjusted) $R^2$, and receiver operator characteristic (ROC) curves.

Oftentimes, the performance of the model is assessed using a different dataset than that which was used to build it, so as to evaluate its wider validity.

### 3.1.4 How is the model interpreted?

The prediction model provides information about the expected value (or risk) of an outcome, given data on the covariates in the model. The model does *not* provide information about how to *change* the expected value (or risk) of an outcome. The consequences of any (hypothetical) intervention to change the outcome may be estimated from external knowledge (e.g. clinical trial results for low-molecular-weight heparin treatment[13]) but are unknowable from the model itself without further strong assumptions.

The model also cannot indicate which *individual* predictors are most relevant, as the *set* of predictors ultimately selected depends upon the initial set of 'candidate' predictors and the parameterisations considered. Moreover, there is no guarantee that the variables which, *as a group*, accurately predict the outcome have any sensible interpretation (causal or otherwise) *in isolation*. In general, even attempting to qualitatively or quantitatively rank the 'contribution' of different predictors should not be attempted, since both the magnitude and sign of each predictor are conditional on the inclusion of all others. While there are a small number of cases where ranking predictors could be useful, it is nevertheless a common problem that authors seek to interpret individual covariates in ways that are inappropriate.[16]

## 3.2 Causal modelling

The goal of causal explanation is to estimate the true causal association[1] between a particular variable (often referred to as the 'exposure') and the outcome, by removing all other hypothesised associations which distort that relationship (henceforth referred to as 'spurious' associations). Such information may then be used to attempt to *alter* the outcome by altering the exposure.

In the context of a GLM, the causal association of interest is represented by the *coefficient* of the exposure variable; removing all spurious associations is achieved in principle by also including as covariates a sufficient set of variables which 'control for' those associations.

Returning to our clinical example, chemotherapy has been identified as a 'risk factor' for VTE.[18] However, determining to what degree it increases the risk of VTE compared to other treatments or interventions (e.g. no chemotherapy) requires a robust estimate of the *causal* effect of chemotherapy on VTE risk.

The *causal* research question (RQ2) can thus be framed as:

---

[1] We restrict our analysis here to considering the 'total' causal effect, since additional complexities arise in the estimation of 'direct' and 'indirect' (i.e. 'mediated') causal effects.[17]

*To what degree does chemotherapy increase the risk of developing VTE?*

### 3.2.1 Which covariates should be considered for inclusion in the model?

Variables which are hypothesised to create or transmit spurious associations between the exposure and outcome should be identified and considered for inclusion in the model. The most familiar of such associations is confounding, which arises due to one or more common causes of the exposure and outcome. The causal effect of chemotherapy on risk of VTE, for example, is likely to be confounded by tumour size, since this is often taken into consideration when deciding whether to initiate chemotherapy and also likely affects subsequent VTE risk.

In causal modelling, it is equally as important to identify variables that should be *excluded* from consideration. For example, spurious associations may arise due to an under-recognised phenomenon known as 'collider bias', in which 'controlling for' a common causal descendent (i.e. a 'collider') induces an additional non-causal dependency between the exposure and outcome.[19, 20] Variables which transmit part of the *causal* association of interest should also be excluded from consideration.

The process of identifying covariates to potentially include or exclude should not be limited by what is available in a particular dataset, since spurious associations between the exposure and outcome do not simply cease to exist if they are ignored.[21, 22]

### 3.2.2 How are covariates selected for inclusion in the model, which covariates are ultimately selected, and how are they parameterised?

The covariates ultimately selected for inclusion in the model must, *as a group*:

1. 'Control for' all spurious associations,
2. Not 'control for' any of the causal association, and
3. Not create any *additional* spurious associations.[23]

Graphical causal models (often in the form of DAGs) are of enormous utility to covariate selection in causal modelling. These models consist of a set of nodes (variables) connected by a set of arrows (representing hypothesised direct causal effects), where an arrow from one variable to another implies that a *change* in the first causes a *change* in the second. Any two variables may also be connected by indirect causal pathways, which are sequences of edges which flow in the same direction, and paths which transmit spurious associations (e.g. confounding paths).[6, 23] A simplified DAG for our example scenario is provided in Figure 1.

*[Insert Figure 1]*

The use of a DAG provides a transparent means of spelling out the causal assumptions underlying a given scenario. Moreover, the subset(s) of covariates which satisfy the three conditions identified previous may be identified algorithmically,[24] since in a DAG framework the three conditions correspond to covariates which, *as a group*:

1. Block all confounding paths,
2. Do not block any causal paths, and
3. Do not open any 'colliding' paths.[23]

For instance, the DAG in Figure 1 implies that age, sex, tumour site, and tumour size should be included as covariates in order to estimate the total causal effect of chemotherapy on VTE risk, since they confound the relationship of interest; platelet count should be excluded, since it mediates the effect of chemotherapy on VTE risk.

If there exist multiple subsets of covariates which satisfy the three conditions, practical considerations may be used to choose between them. For example, subsets containing variables which are not available in the intended dataset, or which are otherwise hard to measure accurately, may be rejected.

Once a suitable set of covariates is identified, the goal of covariate parameterisation is to appropriately modify the exposure-outcome relationship (i.e. remove the spurious component(s) of their association); failure to adequately to 'control for' these covariates may result in residual confounding.[25]

### 3.2.3  How is the model evaluated?

A DAG is, by construction, a map of hypothesised statistical dependencies between variables. Conversely, it implies certain statistical independencies between variables, the existence of which can be tested empirically in the dataset used and then potentially used to further refine the model.[24]

Sensitivity analyses may also be employed to estimate the magnitude of biases arising from unmeasured confounding, residual confounding, or collider bias.[26]

### 3.2.4  How is the model interpreted?

The coefficient for the exposure may be interpreted as an estimate of the total causal effect of the exposure on the outcome, i.e. the total expected change in the value of the outcome that is due to a (potentially hypothetical) change in the value of the exposure. In our example, this corresponds to the expected increase in the risk of VTE that is attributable only to initiation of chemotherapy (i.e. all else 'being equal'[27]). Of course, the validity of this estimate is only as good as the validity of the causal assumptions underlying it.

The model does *not* provide information about the total expected change in the outcome that is due to changes in the *other* model covariates except under extremely restrictive circumstances. In general, estimating the effect of a different 'exposure' requires a different model. Erroneously attempting to interpret multiple coefficients in a single GLM as total causal effects is referred to as the 'Table 2 fallacy'.[28]

## 4 Implications

The distinct goals of prediction and causal explanation result in distinct processes for covariate selection and parameterisation, model evaluation, and model interpretation. For these reasons, GLMs for prediction and causal explanation are not interchangeable and should not be conflated.

Any coefficient in a GLM could potentially represent a true causal effect (either direct, total, or a subset of the total), an association due to uncontrolled confounding or collider bias, or any combination thereof. Interpreting a particular coefficient as an estimate of the total causal effect of that covariate on the outcome requires making the assumption that all other covariates in the model 'control for' all spurious associations, do not 'control for' any of the causal association, and do not create any additional spurious associations. Causal modelling processes have these assumptions explicitly built into their foundations, but prediction modelling processes do not.

The goal of prediction modelling is to develop a useful tool to forecast an outcome that has yet to occur, and so the model-building process is ultimately driven by convenience and other practical considerations. It is well-suited to automated methods for covariate selection and parameterisation, because the specific subset of covariates that is ultimately used to predict the outcome (and the way in which they are parameterised) is relatively unimportant so long as the model has a sufficient degree of internal and external validity.

In contrast, the causal model-building process is necessarily driven by external and *a priori* theory, and thus benefits little from algorithmic modelling methodologies. To estimate the causal effect of one variable on another, one must specify both the possible causal pathways through which those effects are realised and the possible non-causal pathways that transmit spurious associations *before* any modelling is undertaken. Although the process of identifying a suitable subset of covariates which remove all spurious associations between the exposure and outcome may be automated once all causal assumptions are made explicit (often in the form of a DAG), identifying the initial set of variables and specifying the *manner* in which they are likely to transmit spurious associations cannot be automated.

## 4.1 Implications for machine learning

Much of the previous discussion surrounding the application and interpretation of GLMs has direct relevance to machine learning.

Machine learning (ML) refers to the automated, typically algorithmic, detection of meaningful patterns in data, and may thus be viewed as a branch of artificial intelligence.[29] In health research, it is often hailed as the new frontier of data analytics which, combined with big data, will purportedly revolutionise delivery of healthcare (e.g. through 'personalised medicine'), provide new and important insights into disease processes, and ultimately lead to more informed public health policy and clinical decision-making.[30-34] Nevertheless, many ML methods (e.g. neural networks) essentially perform regression, and thus require equally thoughtful implementation.

Whilst the application of ML to health research has potential promise, the distinction between prediction and causation has been largely overlooked in discussions surrounding such promise. Historically, ML methods have been applied to *prediction* tasks, and indeed their most high-profile successes to date involve predictive and/or diagnostic tasks.[35-37] However, as has been demonstrated in the preceding example, prediction modelling requires distinct processes from those required for causal inference. It is not difficult to imagine that it will soon be as simple to perform ML methods in off-the-shelf software as it is to implement GLMs, and that the automation which facilitated such confusion in the realm of GLMs will be played out on an even larger scale in the realm of ML.

Selection of a particular variable into a prediction model does not alone imply anything about the strength (or even existence) of a true causal relationship between that variable and the outcome of interest. More generally, individual 'predictors' offer little insight (causal or otherwise), as prediction models are inherently outcome-focused. Interestingly, complex 'black box' algorithms – which are frequent subjects of ethical concerns[38-41] – are potentially less susceptible to causal (mis)interpretation than GLMs because they do not attach numeric values to particular covariates. However, there is unfortunately already some evidence to suggest that the conflation of prediction and causation has extended into the realm of ML,[42-44] though we have not conducted a systematic review to assess how widespread such behaviour is.

Integration of modern causal inference methods into ML applications should be sought and encouraged for answering causal questions. Indeed, there is already promising research being done in this area (e.g.[45-49]).

# 5   Recommendations

Based upon our previous analysis and discussion, we offer the following three recommendations to ensure that both prediction and causal modelling be used appropriately and to greatest effect in health research.

### *5.1.1   The purpose of any model should be specified from the outset, and the model built with appropriate respect for this context.*

The distinct purposes of prediction and causal inference require distinct models. It is paramount that the purpose of any model is established from the outset, and that it is then constructed, evaluated, and interpreted with appropriate respect for this context. All reporting of methods and results should be consistent with this guidance in order to avoid misinterpretation or misapplication of the model.

### *5.1.2   Contextual knowledge is generally required for both prediction and causal inference, and this cannot be automated.*

The frameworks of both GLMs and ML can provide automated methods for estimating the parameters that map inputs (i.e. covariates) to outputs (i.e. outcomes).[50] These methods do not, however, replace contextual knowledge, which is generally required for both prediction and causal modelling.

Specifying the initial set of variables to consider for each model, for instance, requires temporal knowledge. A model which includes a variable that occurs *after* any (hypothetical) intervention to alter the outcome is of little use practically, yet an algorithm by itself cannot make this determination.

Requirements for contextual knowledge are even greater for causal modelling, where causal assumptions must be specified, ideally before any modelling is undertaken; this is addressed further in point 3.

### *5.1.3   Attributing causal effects requires causal assumptions.*

Robust causal evidence from observational data cannot be obtained in a 'theory-free' environment. Attempting to extract causal meaning from models that have not been built in an explicit causal framework is futile at best. Whilst it might be argued that prediction in one setting helps to inform intervention in another, such transference of inference relies on the ability of a single selected 'predictor' to provide causal insight, which it cannot do without additional strong (*causal*) assumptions.

# 6 Conclusion

The distinction between modelling strategies for prediction and those for causal inference is not widely appreciated in the context of GLMs, despite being the mainstay method for health data analysis. Failing to recognise the distinction and its implications risks wasting substantial financial resources and creating confusion both in academic and public discourse. Moreover, the application of ML and other modelling methodologies are likely to suffer many of the same problems (and potentially on a vastly larger scale) if the lessons of GLMs are not heeded. We hope that this article has highlighted some of the important considerations associated with the use of GLMs and ML for prediction and causal inference, and thereby provides researchers with practical guidance for implementing them.

# 8 Figures

## 8.1 Figure 1

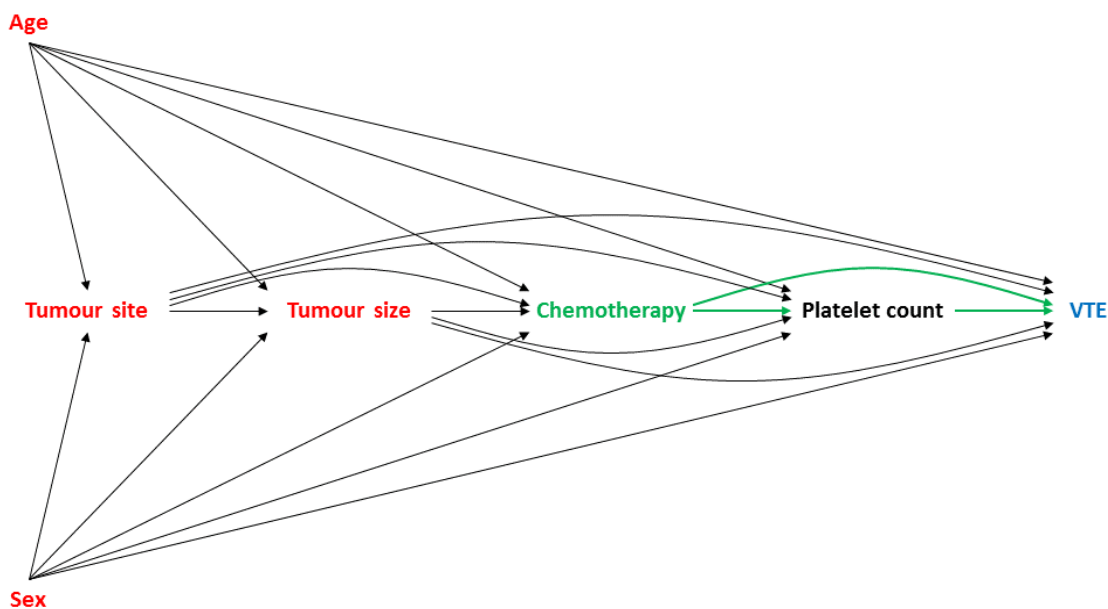

**Figure 1:** Directed acyclic graph (DAG) depicting the hypothesised causal relationship between chemotherapy (the 'exposure', in green) and VTE (the 'outcome', in blue); causal pathways are highlighted in green. Age, sex, tumour site, and tumour size (all in red) confound the relationship between Chemotherapy and VTE, and so should be included as covariates in the GLM in order to estimate the total causal effect of Chemotherapy on VTE risk. Platelet count mediates the relationship between Chemotherapy and VTE, and so should *not* be included as a covariate in the GLM in order to estimate the total causal effect of chemotherapy on VTE risk.

# 9 Sources of funding

This work was supported by the Economic and Social Research Council [grant number ES/J500215/1 to KFA]; The Alan Turing Institute [grant number EP/N510129/1 to PWGT and MSG]; and the Commonwealth Scholarship Commission [to JM].

# 10 Acknowledgments

The Authors would like to thank Professor Kate Tilling (University of Bristol) for her helpful comments on previous versions of this manuscript.